\documentclass[preprint,superscriptaddress,showkeys,showpacs,
tightenlines,fleqn,nofootinbib,prd,byrevtex]{revtex4} 
\usepackage{graphicx}
\usepackage{bm}
\usepackage{epstopdf}
\begin{document}      
\preprint{YITP-08-65}
\title{Scalar susceptibility from the instanton vacuum\\
with meson-loop corrections}      
\author{Seung-il Nam}
\email{sinam@yukawa.kyoto-u.ac.jp}
\affiliation{Yukawa Institute for Theoretical Physics (YITP), Kyoto
University, Kyoto 606-8502, Japan} 
\date{\today}
\begin{abstract}  
The scalar susceptibility ($\chi_s$) of QCD, which represents the response of the chiral condensate to a small perturbation of explicit chiral-symmetry breaking ($m\ne0$), is investigated within the nonlocal chiral quark model (NL$\chi$QM) based on the instanton vacuum configuration for $N_f=2$. We also take into account $1/N_c$ meson-loop (ML) corrections including scalar and pseudoscalar mesons. It turns out that the chiral condensate is modified to a large extend by the ML corrections in the vicinity of $m=0$, whereas its effect becomes weak beyond $m\approx100$ MeV. As numerical results, we find that $\chi_s=-0.34\,\mathrm{GeV}^2$ with the ML corrections and $0.18\,\mathrm{GeV}^2$ without it, respectively. From these observations, we conclude that the ML corrections play an important role in the presence of finite current-quark mass. 
\end{abstract} 
\maketitle
\section{Introduction}
Among various QCD susceptibilities, which associate with the response of the QCD to external sources, scalar (chiral) susceptibility, $\chi_s$ plays an important role in understanding the pattern of explicit chiral-symmetry breaking. One can define it as follows:  
\begin{equation}
\label{eq:SS}
\chi_s=\sum_\mathrm{flavor}
\frac{\partial\langle\bar{q}q\rangle}{\partial m}\Bigg|_{m=0},
\end{equation}
where $\langle\bar{q} q\rangle$ and $m$ stand for the chiral condensate and current-quark mass, respectively. Note that this nonperturbative QCD quantity is also of great importance in various QCD environments, such as the vacuum, finite $\rho$ and/or $T$, and quark-gluon plasma, since it is deeply related to the phase structure of QCD with the physical quark mass. To date, there has been a good amount of related works, for example, QCD (LQCD) simulation~\cite{Karsch:1994hm}, Schwinger-Dyson method~\cite{Min:2008zz,Wu:2008zzb}, multiflavor Schwinger model~\cite{Smilga:1993in,Smilga:1995qf}, linear-$\sigma$ model~\cite{Chanfray:2001tf,Chanfray:2003rs}, Nambu-Jona-Lasinio (NJL)~\cite{Zhao:2006br,Sasaki:2006ww}, perturbative QCD (pQCD)~\cite{Chakraborty:2002yt}, and so on. 

In the present work, we would like to investigate the scalar susceptibility $\chi_s$ in the {\it vacuum} by employing an effective chiral action, derived from the instanton vacuum configuration, {\it i.e.}, the nonlocal chiral quark model (NL$\chi$QM). This highly constrained framework is characterized by inter-instanton distance ($\bar{R}\approx1\,\mathrm{fm}$) and average size of instantons ($\bar{\rho}\approx1/3\,\mathrm{fm}$), which provides a scale of the model, $\Lambda\approx1/\bar{\rho}$. In many practical applications of the model, large $N_c$ and chiral limit have been taken into account, which simplify the usage of the model considerably and has given good descriptions of light-flavor sector~\cite{Diakonov:1985eg,Shuryak:1982qx,Schafer:1996wv}. Attempts to overcome these two limiting cases ($N_c\to\infty$ and $m\to0$) have been done already extensively in Refs.~\cite{Musakhanov:1998wp,Musakhanov:2001pc,Nam:2006ng,Nam:2007fx,Nam:2006sx,Kim:2005jc,Goeke:2007nc}, in which the finite current-quark mass was explicitly treated~\cite{Musakhanov:1998wp,Musakhanov:2001pc,Nam:2006ng,Nam:2007fx,Nam:2006sx}, and meson-loop (ML) corrections, showed a sizable modification to the leading $N_c$ contribution~\cite{Kim:2005jc,Goeke:2007nc}. From these previous works, it also turned out that the ML corrections are crucial  in treating the finite current-quark mass. Hence, it is natural and critical for us to consider the ML corrections carefully in the present work, since we are interested in the response to a finite value of $m$.

We first compute the $m$-dependent constituent-quark mass, $M_0(m)$, by solving saddle-point equations with the ML corrections, scalar and pseudoscalar mesons, generated by bosonizing an instanton-induced effective action for $N_f=2$. We observe that the ML corrections gives considerable modification in $M_0 $ for the region below $m\approx100$ MeV, in comparison to that from the leading-order contribution only. In addition, for computing $\chi_s$, chiral condensate is also computed, resulting in $\langle iq^\dagger q\rangle_{m=150\,\mathrm{MeV}}/\langle iq^\dagger q\rangle_{m=0}\approx0.7$ with the ML corrections, whereas the ratio becomes about $0.4$ without it.  Note that the empirical value for $\langle is^\dagger s\rangle/\langle iu^\dagger u\rangle\sim\langle is^\dagger s\rangle/\langle id^\dagger d\rangle\approx0.8$~\cite{Ioffe:1981kw}. According to Eq~(\ref{eq:SS}), in Minkowski space, we reaches $\chi_s=-0.34\,\mathrm{GeV}^2$ with the ML corrections and $0.18\,\mathrm{GeV}^2$ without it. From these observations, we can conclude that the ML corrections plays an important role in considering small but a non-zero current-quark mass, showing obvious change of the scalar susceptibility in its sign and magnitude. As for the strength of $\chi_s$, since scalar mesons may not be sensitive to a small perturbation of the current-quark mass, the computed values for $\chi_s$ are reasonable in comparison to pseudoscalar susceptibility $\sim1\,\mathrm{GeV}^2$~\cite{Chanfray:2001tf}. 

The present work is organized as follows: In Section II, we briefly 
review the general formalism of the NL$\chi$QM with the ML corrections. In
Section III, we present numerical results and discussions for the chiral condensate and scalar susceptibility. The final Section is devoted to summary and conclusion.

\section{Effective chiral action from the instanton vacuum}
In this Section, we would like to provide a brief introduction of the present theoretical framework. As mentioned in the previous section, we utilize the nonlocal chiral quark model  (NL$\chi$QM), based on the instanton vacuum configuration. Note that all calculations will be performed in Euclidean space hereafter. Otherwise, we will mention. First, we start with a Dirac equation for a quark in the presence of instanton background fields:
\begin{equation}
\label{eq:DE}
\left[i\rlap{/}{\partial}+im-\rlap{/}{A}_{I\bar{I}}\right]\Psi^{(n)}_{I\bar{I}}
=\lambda_n\Psi^{(n)}_{I\bar{I}},
\end{equation}
where $m$ designates the current-quark mass. $A_{I\bar{I}}$ stands for a singular-(anti)instanton solution as follows:
\begin{eqnarray}
\label{eq:IS}
A^{\alpha}_{I\bar{I}\mu}(x)
=\frac{2\bar{\eta}^{\alpha\nu}_{\mu}\bar{\rho}^2x_{\nu}}
{x^2(x^2+\bar{\rho}^2)}.
\end{eqnarray}
Here, $\eta^{\alpha\nu}_{\mu}$ indicates the 't Hooft symbol as given in Ref.~\cite{Diakonov:2002fq}, whereas $\bar{\rho}$ the average size of (anti)instantons. $I$ and $\bar{I}$ stand for instanton and anti-instanton contributions, respectively. If we assume that low-energy nonperturbative QCD properties are dominated by the following fermion (quark) zero-mode ($\lambda_0=0$ generically),
\begin{equation}
\label{eq:DEZM}
\left[i\rlap{/}{\partial}+im-\rlap{/}{A}_{I\bar{I}}\right]\Psi^{(0)}_{I\bar{I}}=0,
\end{equation}
we can write a quark propagator with the zero-mode solution, $\Psi^{(0)}_{I\bar{I}}$, approximately as
\begin{eqnarray}
\label{eq:QP}
S_{I\bar{I}}=\frac{1}{i\rlap{/}{\partial} +im-
\rlap{/}{A}_{I\bar{I}}}  \approx S_0- 
\frac{\Psi^{(0)\dagger}_{I\bar{I}} \Psi^{(0)}_{I\bar{I}}}{im},
\end{eqnarray}
where $S_0$ denotes a free-quark propagator:
\begin{equation}
\label{eq:FQP}
S_0=\frac{1}{i\rlap{/}{\partial}+im}.
\end{equation}
Eq.~(\ref{eq:QP}) can be written alternatively via the Fourier transform of the zero-mode solution:
\begin{equation}
\label{eq:prop2}
S=\frac{1}{i\rlap{/}{\partial}+im+iM(i\partial,m)},
\end{equation}
where momentum and current-quark mass dependent quark mass, $M(i\partial,m)$ is defined by
\begin{equation}
\label{eq:CQM}
M(i\partial,m)=M_0(m)F^2(i\partial).
\end{equation}
Here, $M_0$ means constituent-quark mass. Moreover, since we are interested in finite current-quark mass, $M_0$ has been taken into account as a function of $m$ implicitly. The current-quark mass dependence of $M_0$ will be discussed in detail later. The form factor $F$, which comes from the Fourier transform of the quark zero-mode solution, reads: 
\begin{equation}
\label{eq:FF}
F(t)=2t\left[I_0(t)K_1(t)-I_1(t)K_0(t)-\frac{I_1(t)K_1(t)}{t}\right],
\,\,\,\,t=\frac{|i\partial|\bar{\rho}}{2},
\end{equation}
where $K_n$ and $I_n$ are modified Bessel functions. With all ingredients having considered, one can construct an effective partition function, which brings about the quark propagator given in Eq.~(\ref{eq:QP}), as follows~\cite{Diakonov:2002fq}:
\begin{equation}
\label{eq:PF}
\mathcal{Z}_{\rm eff}=\int{d\lambda}\,{D\psi}\,{D\psi^{\dagger}}
\exp\left[\int d^4x\psi^{\dagger}(i\rlap{/}{\partial}+im)\psi
+\lambda(Y^++Y^-)+N\left(\ln\frac{N}{\lambda
    V\mathcal{M}}-1\right)\right],
\end{equation}
where the $\lambda$ designates the Lagrangian multiplier to make the action exponent~\cite{Musakhanov:1998wp}. Moreover, we have introduced an arbitrary massive parameter $\mathcal{M}$ to make the argument of the logarithm dimensionless and considered the numbers of the instanton and anti-instanton are the same, $N_I=N_{\bar{I}}=N$. Since we are interested in $N_f=2$, the 't Hooft interaction can be written as:
\begin{eqnarray}
\label{eq:YY}
Y^{\pm}_2&=&\int d^4z\,dU\int d^4x_1\,d^4x_2\,d^4x_3\,d^4x_4
\nonumber\\
&\times&
\left[\psi^\dagger(x_1)_{L,R}(i\partial+im)^\mp\Psi^{(0)}_{I\bar{I}}(x_1-z)
\Psi^{(0)\dagger}_{I\bar{I}}(x_2-z)(i\partial+im)^\pm\psi_{L,R}(x_2)\right]
\nonumber\\
&\times&
\left[\psi^\dagger(x_3)_{L,R}(i\partial+im)^\mp\Psi^{(0)}_{I\bar{I}}(x_3-z)
\Psi^{(0)\dagger}_{I\bar{I}}(x_4-z)(i\partial+im)^\pm\psi_{L,R}(x_4)\right].
\end{eqnarray}
Here, the subscripts $L$ and $R$ represent the chirality of quarks and $z$ the instanton center. We introduce a notation for simplicity as $x^\pm=x^{\mu}(\pm i{\bm\sigma},1_{2\times2})$~\cite{Carter:1998ji}. 

An effective action can be obtained from the effective-partition function defined in Eq.~(\ref{eq:PF}). Note that the 't Hooft interaction given in Eq.~(\ref{eq:YY}) can be casted into four-quark instanton-induced interactions by integrating over color orientation $U$. From this Nambu-Jona-Lasinio (NJL) model-like four-quark interactions, one can construct an effective chiral action as a functional of quarks, and scalar and pseudoscalar mesons, $\Phi=(\Phi_1,\Phi_2,\Phi_3,\Phi_4)=(\sigma,\eta,{\bm\sigma},{\bm\phi}$), by an exact bosonization process for $N_f=2$~\cite{Goeke:2007nc,Diakonov:2002fq}.  The meson fields are normalized as $\Phi^2=\sigma^2+\eta^2+{\bm \sigma}^2+{\bm \phi}^2$. Since there must be meson fluctuations around their classical paths, one can consider an additional contribution to the effective chiral action from these fluctuations as meson-loop (ML) corrections. Practically, we can write the effective chiral action with the ML corrections as follows:
\begin{equation}
\label{eq:ECA123}
\mathcal{S}_\mathrm{eff}=\mathcal{S}^\mathrm{LO}_\mathrm{eff}
+\mathcal{S}^\mathrm{NLO}_\mathrm{eff},
\end{equation}
where the leading-order (LO) term is obtained by bosonizing the effective-partition function of Eq.~(\ref{eq:PF})~\cite{Goeke:2007nc}, resulting in a linear-$\sigma$ type action:
\begin{eqnarray}
\label{eq:ECA2}
\mathcal{S}^\mathrm{LO}_\mathrm{eff}&=&-N\ln\frac{N}{\lambda V\mathcal{M}}+1+
2\int d^{4}x\,\left(\sigma^{2}+\eta^{2}+{\bm\sigma}^{2}+{\bm\phi}^{2} \right)
\nonumber\\
&-&\mathrm{Sp}\ln\left[\frac{i\rlap{/}{\partial}+im
+i\tilde{M}_{0}F(i\partial)
\left(\sigma+\gamma_{5}\eta+i{\bm \tau}\cdot{\bm \sigma}
+i\gamma_{5}{\bm \tau}\cdot{\bm \phi} \right)F(i\partial)}
{i\rlap{/}{\partial}+im}\right],
\end{eqnarray}
where $\mathrm{Sp}$ denotes a functional trace, $\int d^4x\,\mathrm{Tr}_{c,f,\gamma}\langle x|\cdots|x\rangle$, in which the subscripts $c$, $f$, and $\gamma$ indicate color, flavor, and Dirac-spin indices, respectively. Note that $\mathcal{S}^\mathrm{LO}_\mathrm{eff}$ satisfies LO saddle-point equations, $\partial\mathcal{S}^\mathrm{LO}_\mathrm{eff}/\partial\Phi_i=0$ for $i=1\sim4$. $\tilde{M}_{0}$ is defined by $\sqrt{\lambda}(2\pi\bar{\rho})^{2}/2N_{c}$ in the large $N_{c}$ limit and equals to $M_{0}/\Phi_{i}$~\cite{Goeke:2007nc}. 

Now, we are in a position to consider the meson fluctuation explicitly, $\Phi\to\Phi+\Phi'$, which corresponds to $1/N_c$ corrections as pointed out in Refs~\cite{Kim:2005jc,Goeke:2007nc}. This next-to-leading-order (NLO) contribution can be derived straightforwardly by functional derivatives of $\mathcal{S}^\mathrm{LO}_\mathrm{eff}$ with respect to the meson fields:
\begin{eqnarray}
\label{eq:ECA3}
\mathcal{S}^\mathrm{NLO}_{\mathrm{eff},i}&=&
\frac{1}{2}\mathrm{Sp}_\Phi\ln\left[\frac{\delta^2
\mathcal{S}^\mathrm{LO}_\mathrm{eff}}
{\delta\Phi_{i}(x)\delta\Phi_{j}(y)} \right]
\nonumber\\
&=&\frac{1}{2}\mathrm{Sp}\ln
\left[4-\mathrm{Tr}
\left[\frac{\tilde{M}_{0}F^{2}(i\partial)}{i\rlap{/}{\partial}+im
+i\tilde{M}_{0}F^{2}(i\partial)\,\Gamma\cdot\Phi}\Gamma_{i}
\frac{\tilde{M}_{0}F^{2}(i\partial)}{i\rlap{/}{\partial}+im
+i\tilde{M}_{0}F^{2}(i\partial)\,\Gamma\cdot\Phi}\Gamma_{i} \right]
\right],
\end{eqnarray}
where $\mathrm{Sp}_\Phi$ runs over the meson fields. Also we define $\Gamma=(1,\gamma_{5},i{\bm\tau},i\gamma_{5}{\bm\tau})$ and $\mathrm{Tr}=\mathrm{Tr}_{c,f,\gamma}$.  Hereafter, we will take into account the finite fluctuation only in the direction of the  isoscalar $\sigma$ field, corresponding to the spontaneous breakdown of chiral symmetry (SB$\chi$S, $\langle\sigma\rangle\ne0$), whereas other fluctuations are set to be zero, resulting in $\Gamma\cdot\Phi\to\sigma$. By combining the LO and NLO contributions, we finally arrive at the following expression for the effective chiral action with the meson-loop (ML) corrections:
\begin{eqnarray}
\label{eq:ECAF1}
\mathcal{S}^\mathrm{LO+NLO}
_{\mathrm{eff}}&=&\underbrace{-N\ln\frac{N}{\lambda V\mathcal{M}}+1
+2\int d^{4}x\,\sigma^{2}
-\mathrm{Sp}_{c,f,\gamma}\ln\left[\frac{i\rlap{/}{\partial}+im
+iM_{0}F^{2}(i\partial)}
{i\rlap{/}{\partial}+im}\right]}_\mathrm{LO}
\nonumber\\
&+&
\underbrace{\sum_{i=1}^{4}\frac{1}{2}\mathrm{Sp}\ln
\left[4-\frac{1}{\sigma^{2}}\mathrm{Tr}
\left[\frac{M_{0}F^{2}(i\partial)}{i\rlap{/}{\partial}
+i\bar{M}(i\partial)}\Gamma_{i}
\frac{M(i\partial)}{i\rlap{/}{\partial}+i\bar{M}(i\partial)}\Gamma_{i} \right]
\right]}_\mathrm{NLO}.
\end{eqnarray}
Here, $\bar{M}(k)$ stands for $m+M_{0}(m)F^{2}(k)=m+M(k)$. It is more convenient to define the effective chiral action in momentum space in order to compute physical quantities:
\begin{eqnarray}
\label{eq:ECA}
\mathcal{S}^\mathrm{LO+NLO}_{\mathrm{eff}}&=&
\mathcal{C}+N\ln\lambda
+2\int d^{4}x\,\sigma^{2}
-V\int\frac{d^4k}{(2\pi)^4}\mathrm{Tr}\ln\left[\frac{\rlap{/}{D}_a}
{\rlap{/}{d}_a}\right]
\nonumber\\
&+&\sum_{i=1}^{4}\frac{V}{2}\int\frac{d^{4}q}{(2\pi)^{4}}
\ln\left[1-\frac{1}{4\sigma^{2}}\int\frac{d^4k}{(2\pi)^4}\mathrm{Tr}
\left[\frac{M_a}{\rlap{/}{D}_a}\Gamma_i
\frac{M_b}{\rlap{/}{D}_b}\Gamma_i\right]\right],
\end{eqnarray}
where $\rlap{/}{D}_{a,b}=\rlap{/}{k}_{a,b}+i\bar{M}_{a,b}$ and $\rlap{/}{d}_a=\rlap{/}{k}+im$ with $k_a=k$ and $k_b=k+q$. The first term, $\mathcal{C}$ represents irrelevant constant terms. By differentiating the effective chiral action with respect to $\lambda$ and $\sigma$, one is led to two-individual saddle-point equations, which satisfy the following vacuum equations
\begin{equation}
\label{eq:variation}
\frac{\partial\mathcal{S}^\mathrm{LO+NLO}_\mathrm{eff}}{\partial\lambda}
=\frac{\partial\mathcal{S}^\mathrm{LO+NLO}_\mathrm{eff}}{\partial\sigma}=0,
\end{equation}
\begin{eqnarray}
\label{eq:nov}
\frac{N}{V}&=&
\underbrace{\frac{1}{2}\int\frac{d^4k}{(2\pi)^4}\,F(k)}_\mathrm{LO}
+\underbrace{\frac{1}{2}\sum^4_{i=1}\int\frac{d^4q}{(2\pi)^4}
\left[\frac{\int\frac{d^4k}{(2\pi)^4}\left[G_i(k,q)-H_i(k,q)\right]}
{\sigma^2-\int\frac{d^4k}{(2\pi)^4}G_i(k,q)}\right]}_\mathrm{NLO},
\\
\label{eq:sigma}
\sigma^2&=&\underbrace{\frac{1}{4}\int\frac{d^4k}{(2\pi)^4}\,F(k)}_\mathrm{LO}
-\underbrace{\frac{1}{4}\sum^4_{i=1}\int\frac{d^4q}{(2\pi)^4}
\left[\frac{\int\frac{d^4k}{(2\pi)^4}\,H_i(k,q)}
{\sigma^2-\int\frac{d^4k}{(2\pi)^4}G_i(k,q)}\right]}_\mathrm{NLO}.
\end{eqnarray}
In the above equations, we have employed the following notations for simplicity:
\begin{equation}
\label{eq:FG}
F(k)=\mathrm{Tr}\left[\frac{iM_a}{\rlap{/}{D}_a}\right],\,\,\,\,
G_i(k,q)=\frac{1}{4}\mathrm{Tr}
\left[\frac{M_a}{\rlap{/}{D}_a}\Gamma_i\frac{M_b}{\rlap{/}{D}_b}\Gamma_i\right],
\,\,\,\,H_i(k,q)=\frac{i}{4}\mathrm{Tr}
\left[\frac{M_a}{\rlap{/}{D}_a}\frac{M_a}{\rlap{/}{D}_a}\Gamma_i
\frac{M_b}{\rlap{/}{D}_b}\Gamma_i\right].
\end{equation}
Detailed evaluations of $F$, $G_i$, and $H_i$ are given in Appendix. By incorporating Eqs.~(\ref{eq:nov}) and (\ref{eq:sigma}), $M_0(m)$ can be calculated numerically. However, instead of doing that, considering the fact that the LO term in Eq.~(\ref{eq:sigma}) is lager than the NLO term by a factor $10\sim20$, we approximate $\sigma^2$ in the denominator of Eqs.~(\ref{eq:nov}) and (\ref{eq:sigma}) as 
\begin{equation}
\label{eq:sigmatilde}
\sigma^2\to\tilde{\sigma}^{2}=\frac{1}{4}\int\frac{d^4k}{(2\pi)^4}\,F(k).
\end{equation}
We note that this replacement also has to do with a proper LO chiral behavior of a meson propagator at small meson momentum ($q\to0$), that is, the Gell-Mann-Oakes-Renner (GOR) relation, as indicated in Refs.~\cite{Diakonov:1985eg,Goeke:2007bj}:
\begin{equation}
\label{eq:mesonpro}
\Pi^{-1}_{\Phi_i}(q)=4\left[\sigma^2-\int\frac{d^4k}{(2\pi)^4}G_i(k,q)\right]
\to\int\frac{d^4k}{(2\pi)^4}\left[F(k)-4G_i(k,q)\right].
\end{equation}
As for the pion ($i=4$) for $N_f=2$ and $m_\pi\approx0.14$ GeV, from Eq.~(\ref{eq:mesonpro}), we obtain the following relation in the leading order $\mathcal{O}(m)$ in Euclidean space:
\begin{equation}
\label{eq:mesonpro2}
\Pi^{-1}_{\pi}(0)=-8\,m\,N_cN_f\int\frac{d^4k}{(2\pi)^4}
\left[\frac{M_a}{k^2+M^2_a} \right]\to
\underbrace{f^2_\pi\,m^2_\pi= 2\,m\,\langle iq^\dagger q\rangle}
_\mathrm{GOR\,relation},
\end{equation}
where we take  $f_\pi\approx93$ MeV, and the pion propagator reads
\begin{equation}
\label{eq:pionpro}
\Pi_\pi(q)=\frac{1}{2f^2_\pi}\frac{1}{[q^2+m^2_\pi]}.
\end{equation}
Although we have used an analytical LO expression for $\langle iq^\dagger q\rangle$ in Eq.~(\ref{eq:mesonpro2}) without explanation, it will be discussed in detail soon. Plugging $\tilde{\sigma}^2$ into Eq.~(\ref{eq:nov}) and using instanton-packing fraction $N/V\approx(0.2\,\mathrm{GeV})^4$~\cite{Diakonov:1985eg}, one can have $M_0$ as a function of $m$. In practical calculations, we make use of the following parameterized form factors, which mimics that in Eq.~(\ref{eq:FF}) qualitatively well, because of numerical simplicity:
\begin{equation}
\label{eq:para}
M(k)=
M_0\left[\frac{2\Lambda^2}{2\Lambda^2+k^2}\right]^2: |k|\le2.0\,\mathrm{GeV},
\,\,\,\,
M(k)=M_0\left[\frac{6}{(|k|\bar{\rho})^3}\right]^2: |k|>2.0\,\mathrm{GeV},
\end{equation}
where we choose $\Lambda\approx1/\bar{\rho}\approx0.6$ GeV. 

\begin{figure}[t]
\begin{tabular}{cc}
\includegraphics[width=7.5cm]{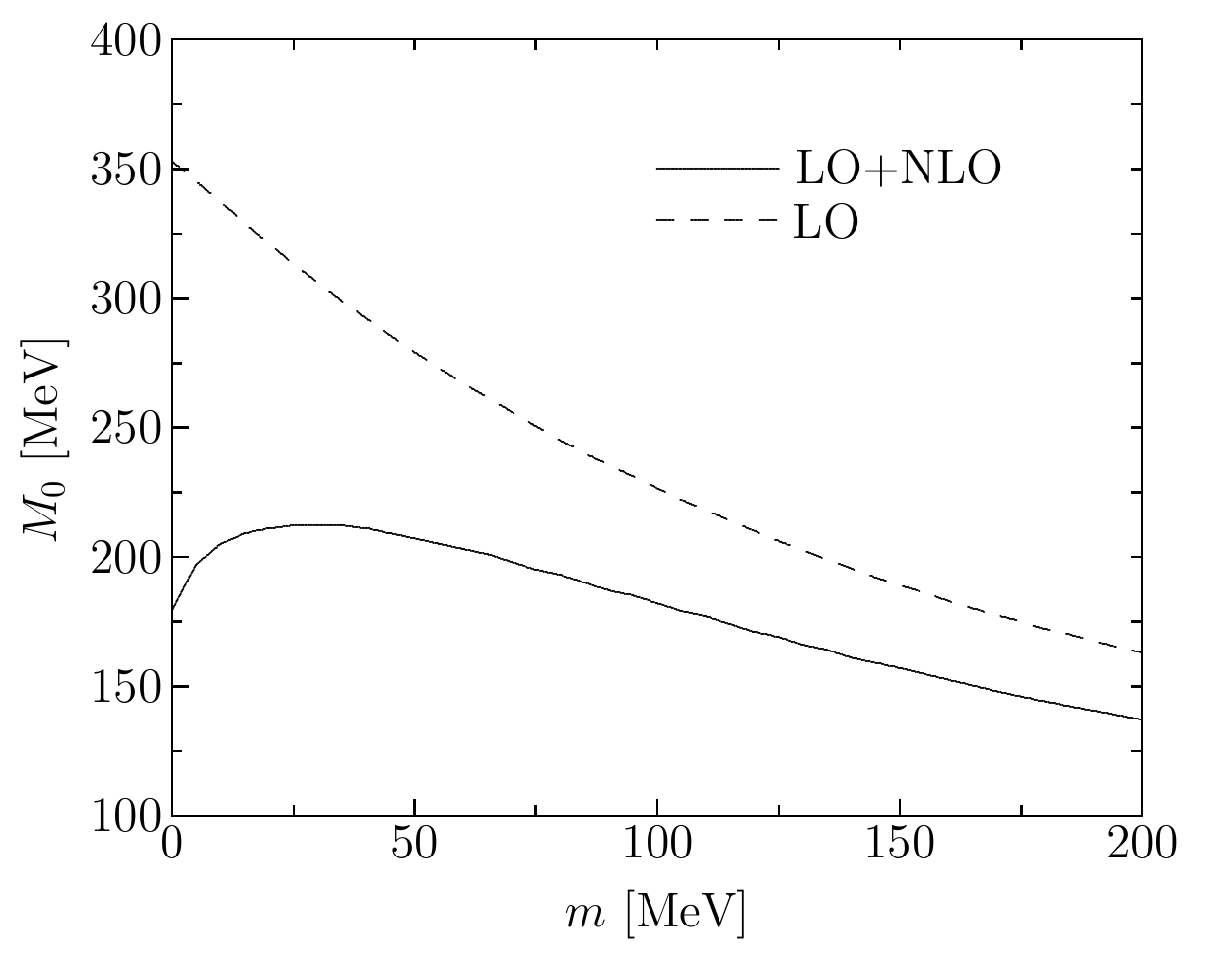}
\includegraphics[width=7.5cm]{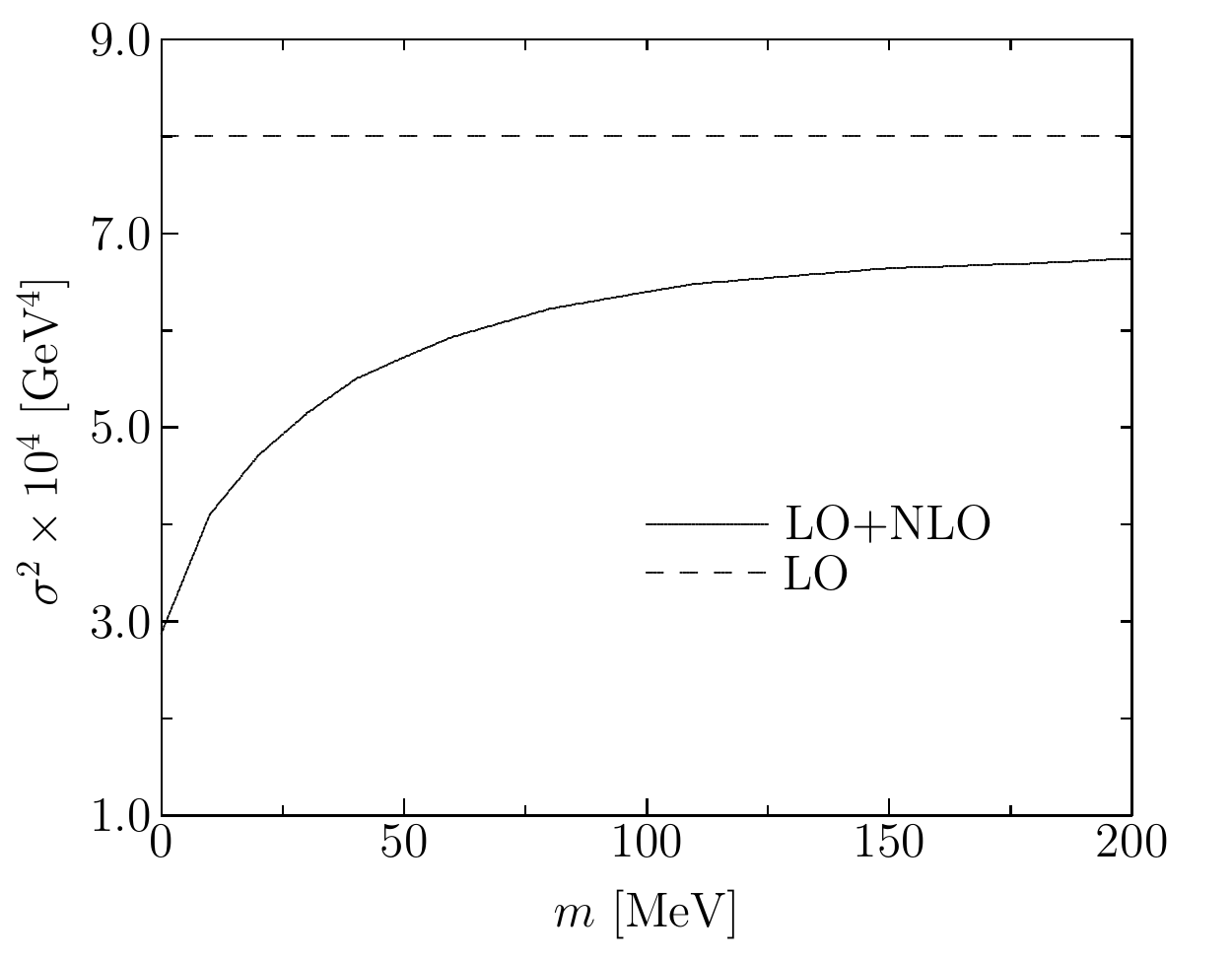}
\end{tabular}
\caption{$M_0$ (left) and $\sigma^2$ (right) as functions of $m$. We draw it for the LO (dashed line) and LO$+$NLO (solid line) contributions, separately.}       
\label{fig1}
\end{figure}

In the left panel of Fig.~\ref{fig1}, we draw numerical results for $M_0$ as a function of $m$ for the LO (without ML corrections) and LO$+$NLO (with ML correction) contributions, separately. It turns out that $M^\mathrm{LO}_0=253$ MeV and $M^\mathrm{LO+NLO}_0=179$ MeV at $m=0$. Being distinctive from $M^\mathrm{LO}_0$, which decreases monotonically with respect to $m$, $M^\mathrm{LO+NLO}_0$ increases and then decreases, indicating the different portion of vacuum effects depending on $m$. In the right panel of Fig.~\ref{fig1}, we depict $\sigma^2$ in the same manner with $M_0$. As for the LO contribution, we have $(\sigma^\mathrm{LO})^2\approx N/(2V)\approx8.0\times10^{-4}\,\mathrm{GeV}^4$. On the contrary, $(\sigma^\mathrm{LO+NLO})^2$ starts from about $3.0\times10^{-4}\,\mathrm{GeV}^4$ and come closer to  $(\sigma^\mathrm{LO})^2$ as $m$ grows. We note that the present results for $M_0$ and $\sigma^2$ with ML corrections are in qualitative agreement with those given in Refs.~\cite{Goeke:2007nc,Kim:2005jc,Goeke:2007bj}, employing the same theoretical framework. In their works, however, $M_0$ and $\sigma^2$ were computed starting from the chiral limit ($m=0$), and the chiral corrections was evaluated. On the contrary, we obtain them directly from the saddle-point equations with the fairly good approximation, $\sigma^2\to\tilde{\sigma}^2$ in the meson propagator, $\Pi_{\Phi_i}$. 

\section{Scalar susceptibility with meson-loop corrections}
\begin{figure}[t]
\includegraphics[width=7.5cm]{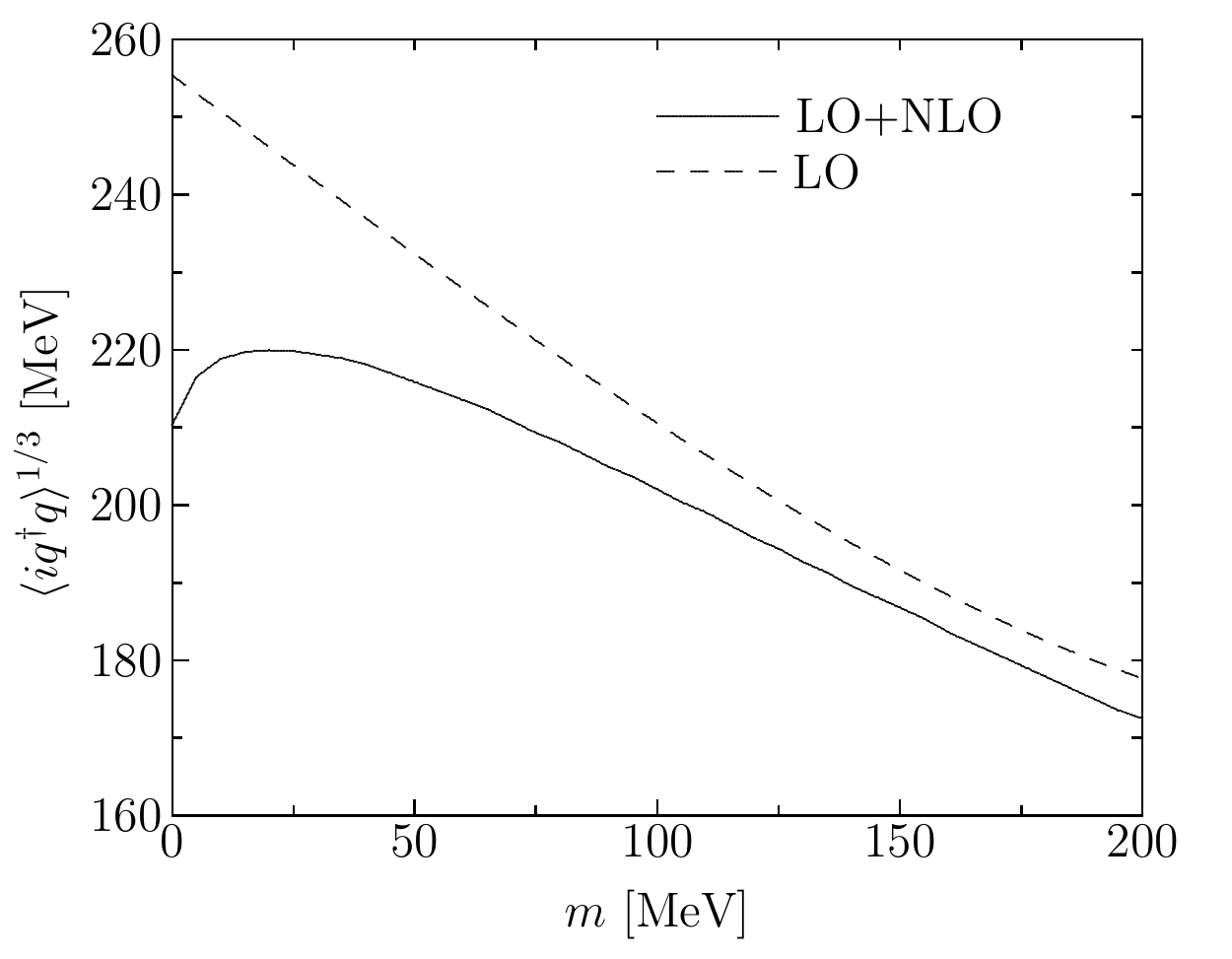}
\caption{$\langle iq^\dagger q\rangle$ as a function of $m$. We draw it for the LO (dashed line) and LO$+$NLO (solid line) contributions, separately.}       
\label{fig2}
\end{figure}

In this section, we will calculate chiral the condensate and scalar susceptibility utilizing the effective chiral action with the ML corrections, derived in Section II. Here, we write the relevant terms of the effective chiral action:
\begin{eqnarray}
\label{eq:ECAprime}
\mathcal{S}^\mathrm{LO+NLO}_{\mathrm{eff}}&=&-V\int\frac{d^4k}{(2\pi)^4}\mathrm{Tr}\ln\left[\frac{\rlap{/}{D}_a}
{\rlap{/}{d}_a}\right]
\nonumber\\
&+&\sum_{i=1}^{4}\frac{V}{2}\int\frac{d^{4}q}{(2\pi)^{4}}
\ln\left[1-\frac{1}{4\sigma^{2}}\int\frac{d^4k}{(2\pi)^4}\mathrm{Tr}
\left[\frac{M_a}{\rlap{/}{D}_a}\Gamma_i
\frac{M_b}{\rlap{/}{D}_b}\Gamma_i\right]\right],
\end{eqnarray}
where $\rlap{/}{d}_a=\rlap{/}{k}_a+im$. The chiral condensate can be derived straightforwardly from $\mathcal{S}^\mathrm{LO+NLO}_\mathrm{eff}$ by a functional derivative with respect to $m$, resulting in:
\begin{eqnarray}
\label{eq:cc}
\langle iq^\dagger q\rangle&=&
\frac{1}{VN_f}\frac{\partial\mathcal{S}^\mathrm{LO+NLO}
_\mathrm{eff}}{\partial m}
\nonumber\\
&=&\underbrace{-4N_c\int\frac{d^4k}{(2\pi)^4}
\left[\frac{\bar{M}_a}{D^2_a}-\frac{m}{d^2_a}\right]
}_\mathrm{LO}+\underbrace{\frac{1}{N_f}\sum^4_{i=1}\int\frac{d^4q}{(2\pi)^4}
\left[\frac{\int\frac{d^4k}{(2\pi)^4}\,H_i(k,q)/M_a}
{\tilde{\sigma}^2-\int\frac{d^4k}{(2\pi)^4}\,G_i(k,q)}\right]}_\mathrm{NLO},
\end{eqnarray}
where $D^2_a=D_aD^\dagger_a$ and $d^2_a=d_ad^\dagger_a$. Note that $\sigma^2$ in the denominator of the NLO term has been replaced by $\tilde{\sigma}^2$ as done previously. In Fig.~\ref{fig2}, we draw the numerical results of $\langle iq^\dagger q\rangle$ for the LO and LO$+$NLO
 contributions, separately. At $m=0$, it  turns out that $\langle iq^\dagger q\rangle^\mathrm{LO}=(253\,\mathrm{MeV})^3$, whereas $\langle iq^\dagger q\rangle^\mathrm{LO+NLO}=(210\,\mathrm{MeV})^3$, showing about $20\%$ difference. Interestingly, $\langle iq^\dagger q\rangle^\mathrm{LO+NLO}$ is saturated once around $m\approx20$ MeV, then decreases, being different from the LO contribution, which decreases monotonically. Two contributions start to behave in a similar manner beyond $m\approx100$ MeV as $M_0$ does. 

Although we have been working with $N_f=2$, it is meaningful to examine the chiral condensate away from $m=0$, to see the effects of the ML corrections for explicit chiral symmetry breaking. Empirically, the ratio between the condensates of light and strange quarks was estimated to be $\langle\bar{s}s\rangle/\langle\bar{u}u\rangle\sim\langle\bar{s}s\rangle/\langle\bar{d}d\rangle=0.8\pm0.3$~\cite{Narison:2002hk} and $0.75\pm0.12$~\cite{Gasser:1984gg}. This ratio in the present framework at $m=0$ and $m=150$ MeV, being computed, it turns out to be $0.71$, which shows relatively good agreement with the empirical values. In contrast, the LO contribution gives the ratio $\sim0.43$. From these observations, we can conclude that the ML corrections improve the model noticeably in the presence of finite current-quark mass. 

Now, we are in a position to compute the scalar susceptibility $\chi_s$, incorporating Eqs.~(\ref{eq:SS}) and (\ref{eq:cc}). In Minkowski space, $\chi_s$ reads:
\begin{eqnarray}
\label{eq:ss}
\chi_s&=&\sum_\mathrm{flavor}
\frac{\partial\langle\bar{q}q\rangle}{\partial m}\Bigg|_{m=0}
=-N_f\frac{\partial\langle iq^\dagger q\rangle}{\partial m}\Bigg|_{m=0}
=\underbrace{4N_cN_f\int\frac{d^4k}{(2\pi)^4}
\left[\frac{D^2_{a0}-2M^2_a}{D^4_{a0}}-\frac{1}{k^2}\right]}_\mathrm{LO}
\nonumber\\
&-&\underbrace{\sum^4_{i=1}\int\frac{d^4q}{(2\pi)^4}
\left[\frac{\int\frac{d^4k}{(2\pi)^4}\left[\frac{1}{M_a}
\frac{\partial H_i(k,q)}{\partial m}\Big|_{m=0}\right]}
{\tilde{\sigma}^2_0-\int\frac{d^4k}{(2\pi)^4}\,G_i(k,q)}
+\frac{\int\frac{d^4k}{(2\pi)^4}\left[\frac{H_i(k,q)}{M_a}
\frac{\partial G_i(k,q)}{\partial m}\Big|_{m=0}\right]}
{\left[\tilde{\sigma}^2_0-\int\frac{d^4k}{(2\pi)^4}\,G_i(k,q)\right]^2}
\right]}_\mathrm{NLO},
\end{eqnarray}
where $\rlap{/}{D}_{a0}=\rlap{/}{k}_a+iM_a$ and $\tilde{\sigma}^2_0=\tilde{\sigma}^2$ at $m=0$. The derivatives of $F$ and $G$ with respect to $m$ are evaluated as:
\begin{eqnarray}
\label{eq:FGdev}
\frac{\partial H_i(k,q)}{\partial m}\Bigg|_{m=0}&=&
\frac{3}{4}\mathrm{Tr}
\left[\frac{M_a}{\rlap{/}{D}_{a0}}\frac{M_a}{\rlap{/}{D}_{a0}}
\frac{1}{\rlap{/}{D}_{a0}}
\Gamma_i
\frac{M_b}{\rlap{/}{D}_{b0}}\Gamma_i\right],
\nonumber\\
\frac{\partial G_i(k,q)}{\partial m}\Bigg|_{m=0}&=&
-\frac{i}{2}\mathrm{Tr}
\left[\frac{M_a}{\rlap{/}{D}_{a0}}\frac{1}{\rlap{/}{D}_{a0}}
\Gamma_i\frac{M_b}{\rlap{/}{D}_{b0}}\Gamma_i\right].
\end{eqnarray}
As for the numerical results, we obtain $\chi^\mathrm{LO}_s\approx0.18\,\mathrm{GeV}^2$ and $\chi^\mathrm{LO+NLO}_s\approx-0.34\,\mathrm{GeV}^2$ in Minkowski space, showing substantial modification from the ML corrections. Especially, the sign difference is crucial due to the obviously different behaviors of $M_0$ near $m=0$ as shown in the left panel of Fig~\ref{fig1}. Note that $\chi_s$ can be estimated from the results of Ref.~\cite{Goeke:2007bj}, giving about $-0.50\,\mathrm{GeV}^2$, which is compatible with ours. Here, an explanation for the strength of $\chi_s$ is given; since scalar mesons must be insensitive to a small perturbation of the current-quark mass being considered, their relatively heavy masses, the computed values for $\chi$ are reasonable in comparison to the pseudoscalar susceptibility, estimated as $\sim1\,\mathrm{GeV}^2$ in Ref.~\cite{Chanfray:2001tf}. All the numerical results up to now are listed in Table.~\ref{table1}.
\begin{table}[t]
\begin{tabular}{c|c|c|c|c|c|c}
&$M_0$ [MeV]
&$\sigma^2$ [GeV$^4$]
&$|\langle iq^\dagger q\rangle|^{\frac{1}{3}}$ [MeV]
&$|\langle iq^\dagger q\rangle'|^{\frac{1}{3}}$ [MeV]
&$\frac{\langle iq^\dagger q\rangle}{\langle iq^\dagger q\rangle'}$
&$\chi_s$ [GeV$^2$]\\
\hline
LO&$353$&$8.00\times10^{-4}$&$255$&192&0.43&$0.18$\\
\hline
LO$+$NLO&$179$&$2.88\times10^{-4}$&$210$&$187$&$0.71$&$-0.34$\\
\end{tabular}
\caption{Various physical quantities computed at $m=0$. Here, we take $m=150$ MeV for $\langle iq^\dagger q\rangle'$.}
\label{table1}
\end{table}
\section{Summary and conclusion}
In the present work, we have investigated the scalar susceptibility of the QCD vacuum, $\chi_s$ and related nonperturbative quantities, such as the chiral condensate $\langle iq^\dagger q\rangle$. For this purpose, we employed an effective chiral action, derived from the instanton vacuum configuration in Euclidean space. Since it has been known that the meson-loop (ML) contribution, corresponding to the large $N_c$ corrections, play important roles in a system beyond the chiral limit and, since we were interested in the response of the chiral condensate to a small perturbation of current-quark mass, we added the ML corrections as a NLO contribution to the LO effective chiral action. By virtue of field-theoretical functional methods, the ML corrections were derived from the LO contribution by fixing meson fluctuations along the direction of the isoscalar $\sigma$ field. 

The saddle-point equations were obtained by differentiating the effective chiral action with respect to the external parameters, $\lambda$ and $\sigma$, giving current-quark mass dependent constituent quark mass $M_0$, self-consistently. Owing to the ML corrections, $M_0$ behaves very differently in the vicinity of $m=0$ from that without the corrections. This difference effected also on the chiral condensate as well as the scalar susceptibility, substantially. Especially, we observed  in Minkowski space that $\chi_s=0.18\,\mathrm{GeV}^2$ without the ML corrections, whereas $\chi_s=-0.34\,\mathrm{GeV}^2$ with it, showing the critical sign difference. We note that this small value, in comparison to that for pseudoscalar susceptibility $\sim1\,\mathrm{GeV}^2$, is moderate, taking into account that heavier scalar mesons might be insensitive to a small perturbation of current-quark mass. 

Consequently, we can verify that the ML corrections are critical in considering a system beyond chiral limit as shown in the drastic change in scalar susceptibility for instance. As a perspective, we are also concerned in the extension of the present results to a system with $N_f=3$, and finite density and/or temperature. These extensions must be important in exploring QCD phase structure and finding QCD critical point as exemplary topics.
\section*{acknowledgment}
The author thanks M.~Oka, who gave a motivation for this work, and appreciates the hospitality during his visiting TITech in Tokyo, Japan. He is grateful especially to H.~-Ch.~Kim and M.~M.~Musakhanov, T.~Kunihiro, A.~Hosaka, and D.~Jido for fruitful discussions on the present work. This work was partially supported by the Grant for Scientific Research (Priority Area No.17070002 and No.20028005) from the Ministry of Education, Culture, Science and Technology (MEXT) of Japan. This work was also done under the Yukawa International Program for Quark-Hadron Sciences. The numerical calculations were carried out on YISUN at YITP in Kyoto University. 
\section*{Appendix}
The functions $F(k)$, $G_i(k,q)$, and $H_i(k,q)$ are evaluated for $i=1\sim4$: 
\begin{eqnarray}
F(k)&=&\frac{4N_cN_fM_a\bar{M}_a}{D^2_a},
\nonumber\\
G_1(k,q)&=&\frac{N_cN_fM_aM_b(k_a\cdot k_b-\bar{M}_a\bar{M}_b)}{D^2_aD^2_b},
\nonumber\\
G_2(k,q)&=&\frac{-N_cN_fM_aM_b(k_a\cdot k_b+\bar{M}_a\bar{M}_b)}{D^2_aD^2_b},
\nonumber\\
G_3(k,q)&=&\frac{-3N_cN_fM_aM_b(k_a\cdot k_b-\bar{M}_a\bar{M}_b)}{D^2_aD^2_b},
\nonumber\\
G_4(k,q)&=&\frac{3N_cN_fM_aM_b(k_a\cdot k_b+\bar{M}_a\bar{M}_b)}{D^2_aD^2_b},
\nonumber\\
H_1(k,q)&=&\frac{N_cN_fM^2_aM_b[\bar{M}_bk^2_a+2\bar{M}_a(k_a\cdot k_b)
-\bar{M}^2_a\bar{M}_b]}{D^4_aD^2_b},
\nonumber\\
H_2(k,q)&=&\frac{N_cN_fM^2_aM_b[\bar{M}_bk^2_a-2\bar{M}_a(k_a\cdot k_b)
-\bar{M}^2_a\bar{M}_b]}{D^4_aD^2_b},
\nonumber\\
H_3(k,q)&=&\frac{-3N_cN_fM^2_aM_b[\bar{M}_bk^2_a+2\bar{M}_a(k_a\cdot k_b)
-\bar{M}^2_a\bar{M}_b]}{D^4_aD^2_b},
\nonumber\\
H_4(k,q)&=&\frac{-3N_cN_fM^2_aM_b[\bar{M}_b
k^2_a-2\bar{M}_a(k_a\cdot k_b)
-\bar{M}^2_a\bar{M}_b]}{D^4_aD^2_b}.
\nonumber
\end{eqnarray}


\end{document}